\documentclass[pre,aps,hidelinks,twocolumn,showpacs,floatfix]{revtex4-1}
\usepackage{lipsum}  
\usepackage{dcolumn}
\usepackage{amsmath}
\usepackage{amssymb}
\usepackage{bm}
\usepackage{graphicx}
\usepackage{epsf,epsfig}
\usepackage{hyperref}
\usepackage{color}
\usepackage{array}
\usepackage{tabularx}
\usepackage{float}
\begin{document}

\title{How cells wrap around virus-like particles using extracellular filamentous protein structures}
\author{Sarthak Gupta$^{1}$, Christian D. Santangelo$^{1}$, Alison E. Patteson$^{1}$, J. M. Schwarz$^{1,2}$}
\affiliation{$^1$ {Physics Department and BioInspired Institute, Syracuse University Syracuse, NY USA}, \\
$^2$ Indian Creek Farm, Ithaca, NY USA}
\date{\today}
\begin{abstract}
 Nanoparticles, such as viruses, can enter cells via endocytosis. During endocytosis, the cell surface wraps around the nanoparticle to effectively eat it. Prior focus has been on how nanoparticle size and shape impacts endocytosis. However, inspired by the noted presence of extracellular vimentin affecting viral and bacteria uptake, as well as the structure of coronaviruses, we construct a computational model in which {\it both} the cell-like construct and the virus-like construct contain filamentous protein structures protruding from their surfaces. We then study the impact of these additional degrees of freedom on viral wrapping. We find that cells with an optimal density of filamentous extracellular components (ECCs) are more likely to be infected as they uptake the virus faster and use relatively less cell surface area per individual virus. At the optimal density, the cell surface folds around the virus, and folds are faster and more efficient at wrapping the virus than crumple-like wrapping. We also find that cell surface bending rigidity helps generate folds, as bending rigidity enhances force transmission across the surface. However, changing other mechanical parameters, such as the stretching stiffness of filamentous ECCs or virus spikes, can drive crumple-like formation of the cell surface. We conclude with the implications of our study on the evolutionary pressures of virus-like particles, with a particular focus on the cellular microenvironment that may include filamentous ECCs.  

\end{abstract}               
\maketitle

%%%FOOTNOTES%%%

\footnotetext{\textit{$^{*}$~E-mail: sgupta14@syr.edu}}
\footnotetext{\textit{$^{**}$~E-mail: csantan@syr.edu}}
\footnotetext{\textit{$^{***}$~E-mail: aepatte02@syr.edu}}
\footnotetext{\textit{$^{****}$~E-mail: jschwarz@physics.syr.edu}}

%Viruses has been part of the evolution on from the beigning of life on earth. Viruses are relatively simpler in stucture, contaning instructions of replication under a membrane with spikes which helps them entering the hosts. They exist at the boundary of dead and living, and only get activated within a host. Viruses has been infected humans by entering in the cell and hijacking the machinary to produce more copies of itself. There are multiple pathways a virus can enter in the cell absorpation, endoctyosis,x,y,z etc. For a virus entering in the cell multiple process should be clicked in unusion i.e. virus should be able to find the appropirate receptor on the cell membrane, it should be in the vicinity of the receptor long enough to form a stable bond then acess one of the entering pathways to enter the membrane. Cell membrane has other biological components on it other than the receptor with which the viurses also intreact and it is not clear if these cell surface components helps or hinders with the endocytosis.

%Previous studies have suggested the possible mechamisim behind endocytosis, but the role of cell surface components are not still clear. In this work, we developed a detailed simulation model of virus with spikes interacting with a cell membrane via cell sufrace components while virus is attached to the receptor. We study how does the density and rigidity of CSV and spikes effect the endocytosis. We also explore the changes in wrapping behaviour of cell membrane due to virus particle rigidity and the stiffness of cell membrane.

\section*{Introduction}

Viruses are relatively simple in structure as they consist of a container and genetic material. The container material ranges from solid-like proteins to fluid-like lipids \cite{Bruinsma2015, Perotti2016}. For the latter type of container, evolutionary pressures have led to the emergence of filamentous spike proteins that protrude out from the container. Such viruses are otherwise known as coronaviruses, examples of which include the common cold, HIV, and, of course, SARS-CoV-1 and SARS-CoV-2 \cite{Einav2019, Ayora-Talavera2018, Laue2021}. The number of spike proteins may indeed vary from virus to virus (for a given type of coronavirus)  \cite{Einav2019, Ayora-Talavera2018, Laue2021}. Moreover, the average spike protein density may change from one type of coronavirus to the next. For instance, the average spike protein density is typically two orders of magnitude in HIV as compared to other coronaviruses \cite{Stano2017a}. Of course, spike protein density is merely one aspect of their characterization. Much work has been done to characterize their binding affinities, and some work has been done to characterize their mechanics \cite{Moreno2022, Kiss2021, Moreira2020, Hu2020, Bosch2003}. In other words, the spike proteins have their own mechanical/conformational degrees of freedom to serve as additional knobs for viruses to optimize their function, which is presumably to replicate.

Since the contents of a virus are minimal, in order to replicate, viruses must enter cells to hijack their biological machinery. Viruses enter cells via multiple pathways. Two of the dominant pathways for coronaviruses are membrane fusion, and endocytosis \cite{Jackson2022a, Grove2011}. In the former, there exists an appropriate receptor on the cell surface, and an additional player arrives just below/at the cell surface to assist the virus in opening its genetic contents. In the latter, the cell surface deforms to wrap around the virus and ultimately pinches it off to enter the cell as a virus-containing vesicle. Clathrin and dynamin are major players in one endocytotic pathway. Other pathways include cytoskeletal filaments, such as actin \cite{Yamauchi2013, Laliberte2011}. Of course, a cell surface contains complex structures with multiple coreceptors interacting with spikes of the virus~\cite{Ripa2021, Maginnis2018}. An identified receptor of the Sars-CoV-2 virus is ACE2; however, other coreceptors such as membrane rafts ~\cite{Ripa2021}, extracellular vimentin \cite{Suprewicz2022, Amraei2022}, glycans~\cite{Hao2021} also play a role in endocytosis in particular. A subset of such coreceptors, such as extracellular vimentin, can be filamentous with their own degrees of freedom~\cite{Ramos2020} and can, therefore, take on a life of their own, if you will, affect viral uptake by the host cell.

Following our initial mostly experimental study of how extracellular vimentin impacts viral uptake \cite{Suprewicz2022}, here, we present a computational model in which {\it both} the cell-like construct and the virus-like construct contain their own filamentous structures protruding from their surfaces and study the impact of these additional degrees of freedom on viral wrapping. Indeed, much research has focused on how the size and shape of the virus impact viral wrapping \cite{Shen2019, Shi2011, Chen2016a} with spherocylindrical objects wrapping (and pinching off) more efficiently than spheres \cite{Vacha2011}. In addition, a recent genetic algorithm approach demonstrated that spheres with patchy sites that are arranged in the lines along the sphere, as opposed to randomly, endocytose more efficiently \cite{Forster2020}. Filamentous structures on viral surfaces have been found to lower energy barriers for binding to the cell \cite{Cao2021a, Stencel-Baerenwald2014}. However, given the presence of filamentous proteins emanating from, or bound to, cell surfaces, how do filamentous objects on the {\it outside} of a cell assist in viral wrapping? 

To begin to answer the above question, we have organized the manuscript as follows. We first present the computational model with several simplifying assumptions. The first is that the cell surface is modeled as a deformable sheet with bending and stretching energy costs, representing the underlying cell cortex supporting the lipid bilayer. Second, as extracellular vimentin is our main filamentous protein candidate, vimentin self-assembles into filamentous, hierarchical structures whose tetramers are six times longer than the thickness of a cell membrane~\cite{ULF2012}. There is some evidence for extracellular vimentin adhering to the cell surface via glycans~\cite{glycans}; however, vimentin may stick directly to the lipid bilayer or to the cell cortex via plectins. Since many of the details regarding how extracellular vimentin interacts with the surfaces of cells have yet to be discovered, we assume that the extracellular filamentous protein structures, or extracellular components, tether directly to the cell surface. Finally, the third simplifying assumption is that while the viral container will contain filamentous spikes emanating from it, the container will be a deformable shell with elastic interactions. Even with such simplifying assumptions, the model is still very rich in its viral wrapping phase space.   We then present the results of the modeling after varying the density of the filamentous, extracellular components, and the filamentous extra viral components, as well as the mechanics of each component and of the deformable sheet, to explore their implications for viral wrapping. We then conclude with a discussion of the evolutionary pressures of viral structure and mechanics, given our results.

%For influenza virus sialic acids (SAs) of the cell surface acts as receptors
%~\cite{Weis1988} and CD14 and TLR acts as coreceptors~\cite{Pauligk2004}.

%with nearest and next-nearest neighbor spring  interactions, with the latter capturing bending energy costs of the deformable sheet. 

%Common viral receptors and co-receptors includes sialyated glycans, immunoglobulin superfamily members, integrins, phosphatidylserine for various virus families ~\cite{Maginnis2018}.

\begin{figure*}[t]
\includegraphics{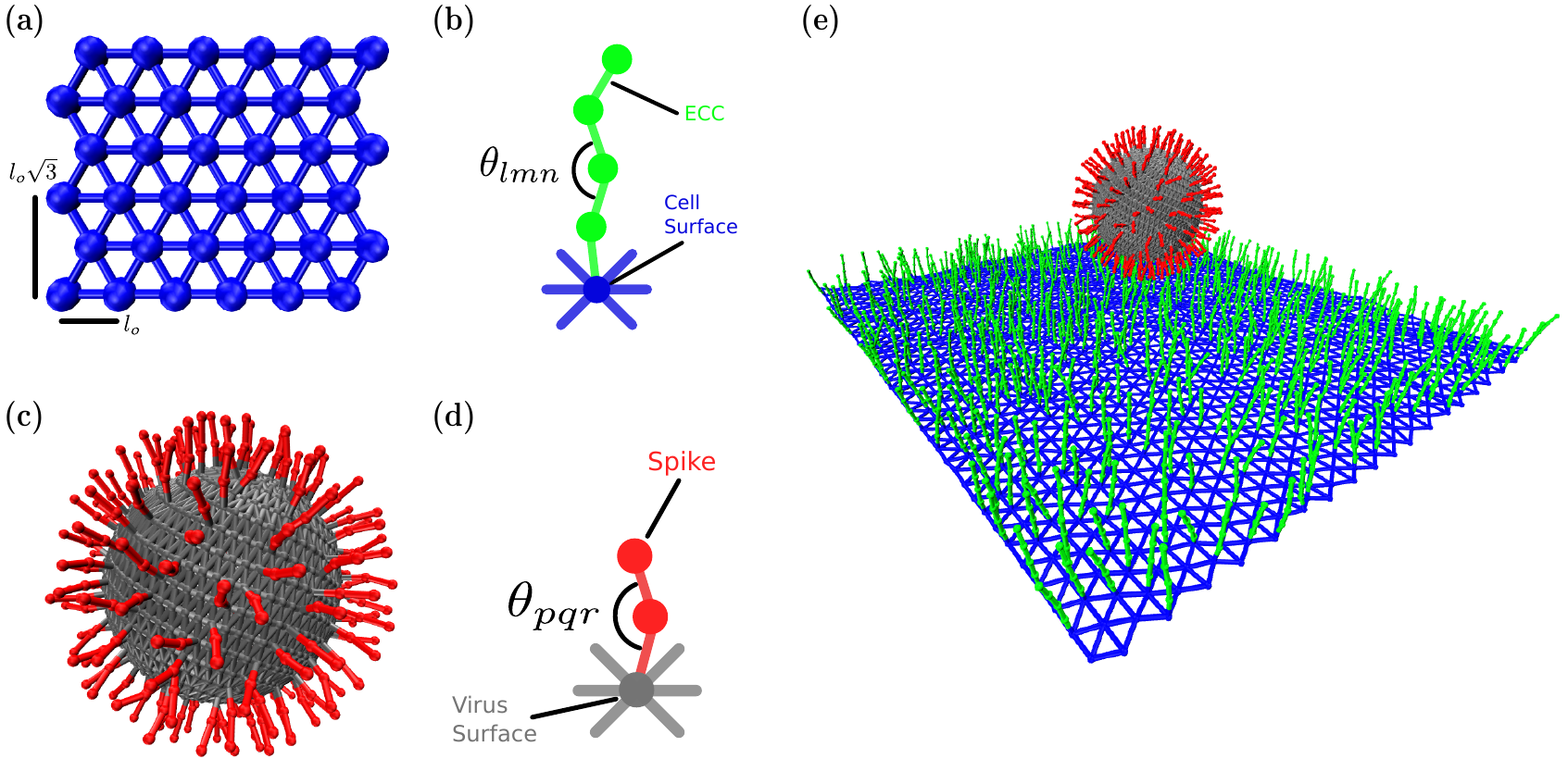}
  \caption{ { \it Schematic illustration of the components of the coarse-grained computational model}  (a) The cell surface is constructed as network of equilateral triangles (blue particles and edges) (b) Extracellular components are modeled as a single polymer with bending and streching rigidity (c) The virus surface (gray) is composed of particles on a sphere connected each other via springs and containing spikes (red) emanating from the surface (d) Spike proteins are quantified as single polymers with bending and stretching rigidity (e) Combining all the components of the computational model yields a virus with spikes interacting with a cell surface with extracellular components. }
\end{figure*}

%Numerical studies suggest [Comparision to various nanoparticle studies]
%\lipsum[2-5]

\section*{Computational model and methods}
To understand the role of the filamentous structures on the outside of both a cell and a virus on viral uptake, we model viral wrapping via coarse-grained Brownian dynamics simulations. Here are the relevant players and how they are incorporated into the model and their respective interactions with the other players.\\

\textit{Cell surface}: We model the cell surface as a deformable sheet, as shown in Fig 1(a). The sheet is made up of particles, shown in blue, that are connected via springs. Nearest-neighbor particles in the sheet interact via harmonic spring potential $V_{Spring}=\frac{K^{Cell Surface}_{NN}}{2}\left ( r_{ij} -l_{o} \right )^{2}$, where $K^{Cell Surface}_{NN}$ is the strength of spring interaction, $r_{ij}$ is the distance between nearest neighbors and $l_{o}$ is the edge length of the equilateral triangle. Bending rigidity is introduced in the surface with a spring interaction between second nearest neighbors $V_{Spring}=\frac{K^{Cell Surface}_{SNN}}{2}\left ( r_{ik} -{l_{o}} \sqrt{3} \right )^{2}$,  where $K^{Cell Surface}_{SNN}$ is the strength of second spring interaction, $r_{ik}$ is the distance between second nearest-neighbors and $l_{o}\sqrt{3}$ is the second nearest-neighbor distance. This second nearest-neighbor interaction acts as a brace to constrain the distance between the two more distant particles such that there is an effective hinge between the two triangles spanning the brace, and given the additional braces, we can explore the bending cost of the deformable sheet. This interpretation is further supported by an additional soft-core repulsion between particles (detailed below) with the particle size relative to the smallest equilateral triangle size set such that the particles cannot pass through this triangle to decrease twisting. Moreover, the presence of the filamentous structures on one side of the sheet also helps to minimize twisting.

\begin{table}[H]
\centering

\begin{tabularx}{0.5\textwidth} { 
   >{\raggedright\arraybackslash}X 
   >{\centering\arraybackslash}X 
   >{\raggedleft\arraybackslash}X  }
 \hline\\
 Parameters & Value & Reference \\\\
\hline
\hline\\
 $K_{NN}^{Cell Surface}$  & $ 1\,\,pN/nm$ &   \\\\
% $K_{SNN}^{Cell Surface}$ & $ 0/0.05/0.1/0.15/0.2\,\, pN/nm$ &   \cite{Eid2020}  \cite{Dimova2014} \cite{Bermudez2004}\\\\
 $K_{SNN}^{Cell Surface}$ & $ 0/3.5/7/10.5/14\,\,  k_{B}T$ &   \\\\
 $K_{Spring}^{Virus}$     & $ 0.05/0.5/1/5/10/50\,\,pN/nm  $  & \cite{Kiss2021} \cite{Zeng2017}\cite{DePablo2019}  \\\\
 %, Brone mosaic virus \cite{Zeng2017} ,influenza virus\cite{Schaap2012} , Human Adenovirus \cite{DePablo2019} Deformable NP\cite{Chen2018a} \\\\

 $K_{Spring}^{Spike}$     & $ 10^{-2}/10^{-1}/10^{0}/10^{1}/10^{2}\,\,pN/nm $ & \cite{Moreira2020}\\\\
 $K_{Spring}^{ECC}$  & $ 10^{-2}/10^{-1}/10^{0}/10^{1}/10^{2}\,\,pN/nm $ &\cite{Schepers2020rr}\\\\
  
 $K_{Spring}^{Receptor}$      & $ 50\,\,pN/nm $& \cite{Cao2021}\\\\  
 $K_{Spring}^{ECC-Receptor}$& $ 5\,\,pN/nm  $& \cite{Bai2020}\\\\

$K_{Bending}^{Receptor}$& $ 120\,\,k_{B}T  $& -\\\\
$K_{Bending}^{ECC}$& $ 24\,\,k_{B}T \times 10^{-2}/10^{-1}/10^{0}/10^{1}/10^{2}   $&- \\\\
$K_{Bending}^{Spike}$& $ 24\,\,k_{B}T \times 10^{-2}/10^{-1}/10^{0}/10^{1}/10^{2}  $& -\\\\
%$D_{T}^{Virus}$& $ 10^{-2} nm^{2}/ms  $& \cite{Kiss2021}\\\\

${\epsilon}_{Attractive}^{ECC-Spike}$  & $ 25\,\,k_{B}T $& -\\\\
$K_{Soft-Repulsion}$  & $ 1\,\,pN/nm $& -\\\\
$D$  & $ 1\,\,\mu m^{2}/s $& -\\\\
 \hline
\end{tabularx}
\caption{Table of parameters used unless otherwise specified. }
\label{table:1}
\end{table}

Again, we are modeling the cell surface as a deformable sheet. With this assumption, we can also explore local, nontrivial shapes of the underlying cortex, which, in turn, drives the plasma membrane shape. On the other hand, since earlier work demonstrates that the head domain of vimentin can also associate with a plasma membrane~\cite{plasmamembrane}, we will address the potential changes if the sheet were to be fluid-like in the discussion. \\

\textit{Extra-Cellular Components (ECC)}: The filamentous ECCs are semiflexible polymers also modeled as particles connected with springs, as shown in Fig 1(b). Each ECC consists of four particles, with the first one connected to a cell surface particle, as shown in Fig. 1(e). Since the cell surface microenvironment may be random, as opposed to a pattern, we place the ECC randomly on the deformable sheet. While a major candidate ECC we consider is extracellular vimentin, given its increasingly pronounced role in viral infection \cite{Ramos2020}, one can also consider heparan sulfate~\cite{Zhang2020}, proteoglycans, glycolipids \cite{Marsh2006}, or sialic acids \cite{Sieczkarski2002}. Each of the candidates may have different mechanical properties based on the protein's physical properties and cell type. To model such properties, each of these co-receptors has bending rigidity given by $V_{Bending}=\frac{K_{Bending}^{ECC}}{2}\left ( cos (\theta_{lmn})-1 \right )^{2}$ where $\theta_{lmn}$ is the angle between three consecutive particles in co-receptors and $K_{Bending}^{ECC}$ is the strength of the bending force. We maintain the equilibration angle to be zero or a straight polymer, such that there is an energy cost for bending. The stretching energetic cost is governed by $V_{Spring}=\frac{K_{Spring}^{ECC}}{2}\left ( r_{ij} -\sigma_{o} \right )^{2}$, where $K_{Spring}^{ECC}$ is the strength of spring interaction, $r_{ij}$ is the distance between nearest neighbor and $\sigma_{o}$ is the diameter of a ECC particle. Finally, we assume that these filamentous extracellular components are bound to the cell cortex just beneath the lipid bilayer \cite{Pinto2022, Liu2021b, Zhang2020}. \\

\textit{Receptor}: A short receptor is placed in the middle of the cell surface and is the same as the ECC receptor with different values for $K_{Spring}^{Receptor}$ and $K_{Bending}^{Receptor}$ as specified in Table 1. The initial condition is the virus-like particle bound to the short receptor.

\textit{Virus Surface}: The deformable virus surface is initialized by generating particles on a sphere. The particles are arranged in a Fibonacci sequence. We then implement a Delaunay triangulation to identify the edges between these particles. All particles on the sphere are connected with a harmonic spring potential $V_{Spring}=\frac{K_{Spring}^{Virus}}{2}\left ( r_{ij} -r_{o} \right )^{2}$, where $K_{Spring}^{Virus}$ is the strength of the spring and $r_{o}$ is the equilibrium distance found by the triangulation. This construction of a deformable, elastic virus surface, with ultimately filamentous structures emanating from it, is a simplifying assumption as viruses with spike proteins typically consist of fluid-like containers but, nonetheless, provides a starting point for the modeling.

\textit{Spikes on Virus:} Spikes are two particles joined by harmonic springs, as shown in Fig. 1(d), which are also joined to the virus surface by another harmonic interaction, as shown in Fig 1(c). Since spikes can take on multiple configurations\cite{Turonova2020a, Romer2021}, they also have bending potential with $V_{Bending}=\frac{K_{Bending}^{Spike}}{2}\left ( cos (\theta_{pqr})-1 \right )^{2}$. The connecting spring potential is given by $V_{Spring}=\frac{K_{Spring}^{Spike}}{2}\left ( r_{ij} -r_{o} \right )^{2}$, where $K_{Spring}^{Spike}$ is the strength of the spring and $\sigma_{o}$ is the particle's diameter. We place the spikes on the virus's surface such that the typical distance between two neighboring spikes is between 3-15 nm, the range found in experiments \cite{Kiss2021, Yao2020}.\\

\textit{Spike-receptor interaction}: The virus is connected to the primary receptor via a  harmonic spring interaction, $V_{Spring}=\frac{K_{spring}^{Spike-Receptor}}{2}\left ( r_{ij} -\sigma_{o} \right )^{2}$,  where $K_{spring}^{Spike-Receptor}$ is the strength of the spring and $r_{o}$ is the equilibrium distance. The receptor is made up of four particles connected via harmonic spring and has some bending rigidity.\\

\textit{Spike-ECC interaction}:  Virus spikes and extracellular components may interact via specific binding domains and charge interactions \cite{Stencel-Baerenwald2014, Nguyen2020, Zhang2020}. We model all the possible interactions with a simple, attractive potential, given by Eq. 1, which attracts in some range with the strength of $\epsilon_{Attractive}^{ECC-Spike}$. Although the virus spikes may have multiple receptor binding domains \cite{Lan2020, Maginnis2018}, we took a conservative approach and only allowed one ECC to bind to one spike protein for modeling lock and key interactions \cite{Powezka2020}. Thus, our sticky potential only attracts the top ECC particle and top spike particles to each other. Therefore, $r_{ij}$ is the distance between $ith$ top spike particle and the $jth$ top ECC particle, and $\sigma_{o}$ is the diameter of the particle, or

\begin{equation}
   V_{LJ}=
    \begin{cases}
      4 \epsilon_{Attractive}^{ECC-Spike} \left [ \left ( \frac{\sigma_{o}}{r_{ij}} \right )^{12} - \left ( \frac{\sigma_{o}}{r_{ij}} \right )^{6} \right ] & r\leqslant 2\sigma_{o}\\
      0 &  r> 2\sigma_{o}.
    \end{cases}       
\end{equation}

{\textit Soft-core repulsion}: All components of the model have volume exclusion due to soft-core repulsion given by Eq. 2, where $K_{Soft-Repulsion}$ is the strength of the soft repulsion, $r_{ij}$ is the distance between the centers of two particles, and $\sigma_{o}$ is the diameter of a particle. The repulsion force acts only when the distance between particles is lower than the diameter of the particle, or 

\begin{equation}\
   V_{Soft-Repulsion}=
    \begin{cases}
      \frac{K_{Soft-Repulsion}}{2}\left ( r_{ij}-\sigma_{o} \right )^{2} & r_{ij}\leqslant \sigma_{o}\\
      0 &  r_{ij}> \sigma_{o}.
    \end{cases}       
\end{equation}

\textit{Method:} We implement Brownian dynamics to quantify the dynamics of the model. The equation of motion is given by Eq. 3 below, where $r_{i}$ is the position of an ith particle, $F_{i}^{c}$ is the sum of conservative forces acting on the ith particle, and $\xi(t)$ is the Gaussian white noise to simulate thermal fluctuations and follows $<\xi(t)>=0$ and $<\xi_{i}(t)\xi_{j}(t')>=0$. The sum of conservative forces on the $ith$ particle is $F_{i}^{c}=F_{i}^{Spring}+F_{i}^{Bending}+F_{i}^{LJ}+F_{i}^{Soft-Repulsion}$. We obtain these forces by taking derivatives of the above potentials with respect to the $ith$ particle coordinate ($F_{i}=- \frac{\partial V }{\partial r_{i}} $). We integrate the equation of motion by the Euler-Murayama method, or

\begin{equation} 
\dot{r_{i}}= \mu F_{i}^{c} + \sqrt{2D\xi(t)}.
\end{equation}

\begin{figure*}[t]
\includegraphics{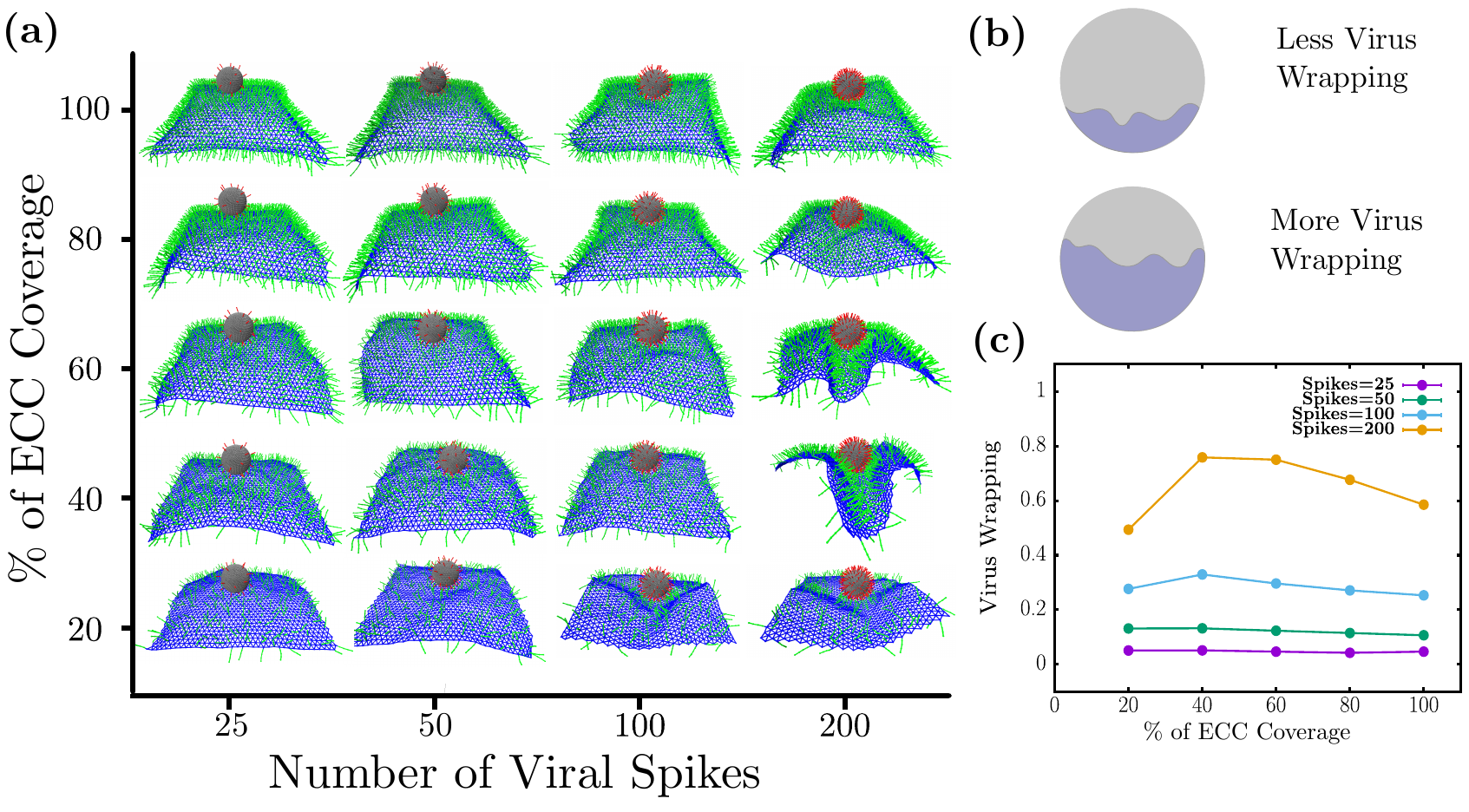}
  \caption{ {\it Optimal coverage of extra-cellular components (co-receptors) is required for maximal wrapping} (a)  Simulation snapshots the virus wrapping as a function of number of virus spikes and ECC coreceptor percent of area coverage.  (b) Virus wrapping is defined as the ratio of virus surface area covered by cell surface and the entire virus surface area (c) Virus wrapping as a function of percent of ECC coverage: For lower spike numbers, there is no appreciable wrapping, but for 200 spikes, non-monotonic viral wrapping behaviour emerges. Note that 200 spikes correspond to a spike area density of 6.3 $\times 10^{-3}$ nm$^{-2}$ and 100 $\%$  ECC coverage corresponds to the area density of 5.4 $\times 10^{-3}$ nm$^{-2}$. }
\end{figure*}

\textit{Scales:} 
Our Brownian dynamics simulation is a coarse-grained simulation. All simulation quantities are normalized via length, time, and force scales. We can convert simulation quantities to biologically relevant quantities by the following definitions: one unit simulation length is defined as 10 nm, one unit simulation time is 1$\mu$s, and one force unit is $10^{-1} $pN. All simulation quantities are expressed in terms of these basic units. We run the total simulation with dt $10^{-4}$ for $10^{8}$ simulation steps or 50 ms and recording positions at each 25 $\mu$s. Total run time is comparable to the typical viral endocytosis time\cite{Imoto2022, Chanaday2018}. To find the optimal conditions for endocytosis in our system, we vary the density of ECC and virus spike, spring strength and bending rigidity of ECC and virus spike, and spring strength of virus. Finally, the bending rigidity of the cell surface can be written in terms of $k_{B}T$. Since bending rigidity is encoded in the second nearest-neighbor springs, multiplying that spring constant with the area of the triangle made up of those springs, with side $l_{o}\sqrt{3}$, gives the bending rigidity. Converting to a dimensionless bending rigidity,  $\tilde{K}_{SNN}^{Cell Surface}$, using our normalized length and force scales, we get that $K_{bending}^{Cell Surface}$=$\tilde{K}_{SNN}^{Cell Surface} \times 0.7 k_{B}T$ gives the values shown in Table 1 to give a comparison with measured bending rigidities of the cell membrane that includes the underlying cortical cortex.\\

The simulation box contains up to 8624 particles. Each particle has a diameter of 10 nm. Since the typical size of the virus is 10s-100s of nanometers and the cell size is usually at the scales of 10 s of micrometers, thus virus interacts only a small part of the cell surface. So, we simulated only a small patch of the cell surface of size 550 nm$\times$480.6 nm, which is made up of 1444 particles. The cell surface boundary is free in our simulations. Each ECC has a length of 40 nm, which is in the range of many cell surface proteins \cite{Patteson2020}. Since ECC density, or percentage of coverage, is a parameter, total ECC particles vary from 1156 to 5776. The primary receptor has four particles as well. The virus surface has a typical diameter of 100 nm~\cite{Sahin2020, Ke2020} and is made up of 1000 particles. The virus spikes consist of two particles or 20 nm long, which correspond to many virus types~\cite{Kiss2021, Bosch2003}. The total number of spike particles varies from 50 to 400. The virus spikes and ECCs attract each other only if the distance between their top particles is less than 8.7 nm which is the range of interaction found in experiments~\cite{Xie2020}. We computed ten realizations for each parameter set and averaged them for plotting purposes. Error bars are the standard deviation of the mean.

\begin{figure*}[t]
\includegraphics{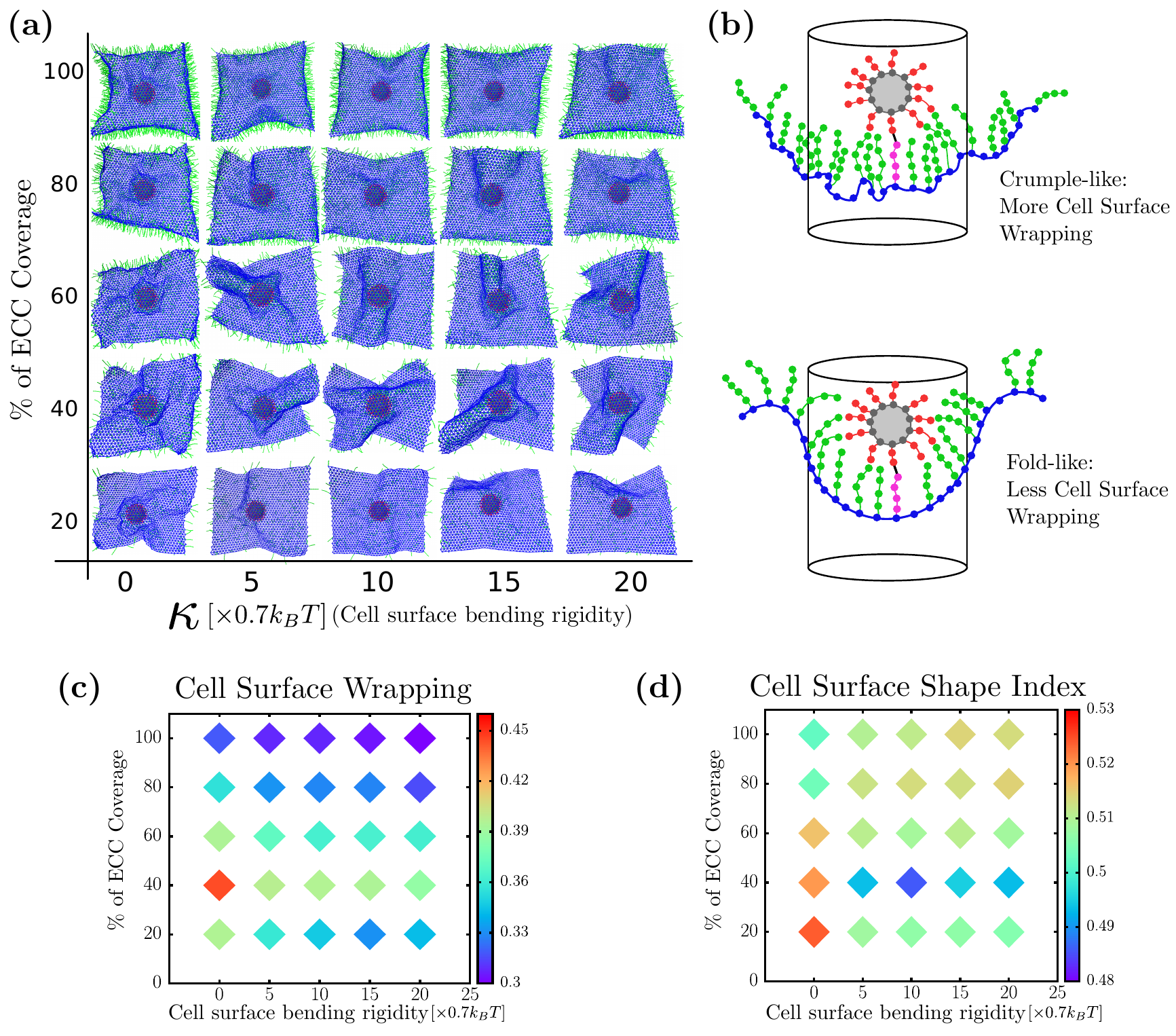}
  \caption{{\it Cell surface rigidity helps generate folds} (a) Bottom view of virus wrapping by the cell surface 
(b) Cell surface wrapping is defined by determining the ratio of the cell surface area inside the cylinder to the total cell surface area. Crumple-like wrapping typically leads to more cell surface area inside the cylinder than fold-like wrapping shape. (c) Cell surface wrapping is minimum for an optimal coreceptor coverage, or area density, which corresponds to folds. Folds use relatively less cell surface area and maximize viral wrapping. (d) Shape index of the cell surface shows folds have a higher value than crumples. Folds form at optimal cell surface coreceptor percent coverage with non-zero cell surface rigidity. }

\end{figure*}

\section*{Results}
\subsection{Optimal co-receptor percentage of coverage yields maximum wrapping}
We first investigate the effects of varying extracellular components (ECC) and the virus spike density on wrapping. We altered the cell surface coverage by the ECCs from 20 percent to cover it fully and varied the spike numbers from 25 to 200 spikes. In Fig2A, we show a typical side view of the final wrapping configurations of the system as the number of virus spikes (horizontal axis) and the percentage of coverage of extracellular components (vertical axis). We observe that the cell surface wraps poorly for lower spike numbers. In other words, there is little interaction between the virus and the cell. However, for 200 spikes, there is substantial wrapping, so we will focus, for the most part, on this part of the parameter space for the number of spikes. Note that 200 spikes uniformly distributed on the surface of the virus-like particle of radius 50 nm leads to an approximate area density of 6.3 $\times 10^{-3}$ nm$^{-2}$. ECC coverage of 100 $\%$ corresponds to the area density of 5.4 $\times 10^{-3}$ nm$^{-2}$.

To quantify the wrapping behavior, we define virus wrapping as the ratio of the surface area covered via a cell surface divided by the whole surface area of the virus, as shown in Fig 2B. To find the surface area covered by the cell surface, we compute all virus spikes adhering to ECC and then add their patch area on the virus surface to obtain the area covered. The fractional area of the virus is what is plotted with unity, denoting the entire viral surface is bound to the cell surface. From Fig 2C, we observe that viral wrapping does not show appreciable changes for viruses with a lower spike number than 200. Given that only one-to-one interaction is allowed between spikes and ECCs, this indicates that only having enough ECC does not necessarily ensure wrapping by the cell surface. However, viruses should also have enough spikes to attach to the cell surface.

We find non-monotonic behavior for the wrapping as a function of the percentage of coverage of extracellular components for 200 spikes, as shown in Fig. 2C. Specifically, there is less virus wrapping at low ECC percent coverage, $>20$ $\%$. Viral wrapping increases at medium percent coverage, 40 and 60 $\%$, only to again decrease for the high ECC percent coverage of 80 and 100 $\%$. Since we have the constraint that a spike can only adhere to one ECC at a time, getting low virus wrapping with low ECC coverage is somewhat expected. As there are few coreceptors to stick to, such viruses cannot interact with many coreceptors. On the other hand, for higher percent coverages of coreceptors, there is a shielding effect that reduces the viral wrapping. Once the coreceptors near the virus attach to the spikes, far away coreceptors cannot interact with virus spikes due to volume exclusion. Thus, having many coreceptors does {\it not} lead to higher virus wrapping. Virus wrapping is maximum at the medium coreceptors density, where virus spikes have enough coreceptors to interact with those around them and have enough space between spikes not to invoke shielding effects. Thus, we found an optimal percent coverage of cell surface coreceptors for maximal wrapping. The notion of an optimal percent coverage for maximal wrapping is a rather reasonable one if one considers effective cell surface bending rigidity that depends on such a quantity. More specifically, as the ECC percent coverage increases, the sheet stiffens such that the sheet eventually can no longer wrap around the viral. At lower ECC percent coverage, some stiffening of the sheet gives rise to a more coordinated wrapping, which we explore in more detail in the next subsection.

\subsection{Cell surface rigidity can drive folds}

Next, we investigate the cell surface bending rigidity's role in viral wrapping. Since there exists heterogeneity in the structure and mechanics of the cell cortex, as viruses invade cells, they may encounter different cellular surface rigidities, which may affect the rate of its uptake. In Fig. 3a, we plot the percent coverage of ECCs on the vertical axis and cell surface bending rigidity on the horizontal axis, showing the typical morphology the cell surface takes during viral uptake. These simulation snapshots are taken from the inside of the cell surface with the virus on the outside of the cell. At lower and mid-ECC percent coverage, we observe that the cell surface exhibits crumples, with more cell surface undulations, for zero bending rigidity. However, we observe fold-like structures of the cell surface for non-zero bending rigidity. Furthermore, we do not observe a significant change in cell surface morphology at high ECC percent coverage, irrespective of cell surface bending rigidity, as the ECC percent coverage presumably dominates the effective bending rigidity.

Given that we have two general types of wrapping configurations that the cell surface takes on, crumple-like wrapping and fold-like wrapping, how do we distinguish between the two types? We say ``crumple-like'' because the crumpling is very localized near the virus, unlike the crumpling of a sheet of paper, for example~\cite{crumpling}. Quantifying folds versus crumples for a triangulated mesh is typically an exercise in discrete geometry. Generically, folds have fewer changes in sign of the local Gaussian curvature than crumples, as the latter consists of shorter, randomly oriented creases. Edge effects combined with noise in the local Gaussian curvature in our system make it difficult to quantitatively distinguish between the two types~\cite{Magid}. So we, instead, measure a dimensionless area of the sheet, much like a gyrification index (GI) as well as a cell surface shape index. As for the former, for a given cylindrical area just beneath the virus, a smoother surface has a smaller area indicating folds, rather than a rougher surface, or larger area, indicating crumples, assuming similar heights. We calculate the cell surface wrapping by taking a circle of radius $R$, centered at the center of mass of the virus. Starting with the highest spike on the virus that is adhered to an ECC, we construct a cylinder around the virus and identify all the cell surface particles in this cylinder for a given radius. As these particles are a part of a triangulated lattice, we see the total surface area of the cell surface inside the cylinder by adding up all the triangle areas inside the cylinder. Here, the radius of cylinder $R$ sets how close or far from the virus we find the surface area. To arrive at a dimensionless area, we divide the obtained surface area by the total area of the sheet. We cannot have a large value of $R$ compared to the virus radius as there is a finite amount of cell surface; by taking too small $R$, on the other hand, we may miss some of the data. Therefore, $R$ is 1.5 times of radius of the virus or 150 nm. For other radii, we studied the quantification, and it is robust for a range of $R$. Please see SI Fig. S1. For calculating the cell surface shape index, we trace the boundary of the cell surface inside the cylinder and determine the perimeter $P$ as well as the cell surface area $A$ inside the cylinder. We then define a dimensionless quantity called shape index as $\sqrt{A}/P$. With this definition of shape index, a hemispherical surface has a shape index of approximately $0.399$, which would be a lower bound. Larger shape indices denote less spherically symmetric wrapping and, so, less efficient wrapping in terms of the use of cell surface area.

\begin{figure*}[t]
\includegraphics{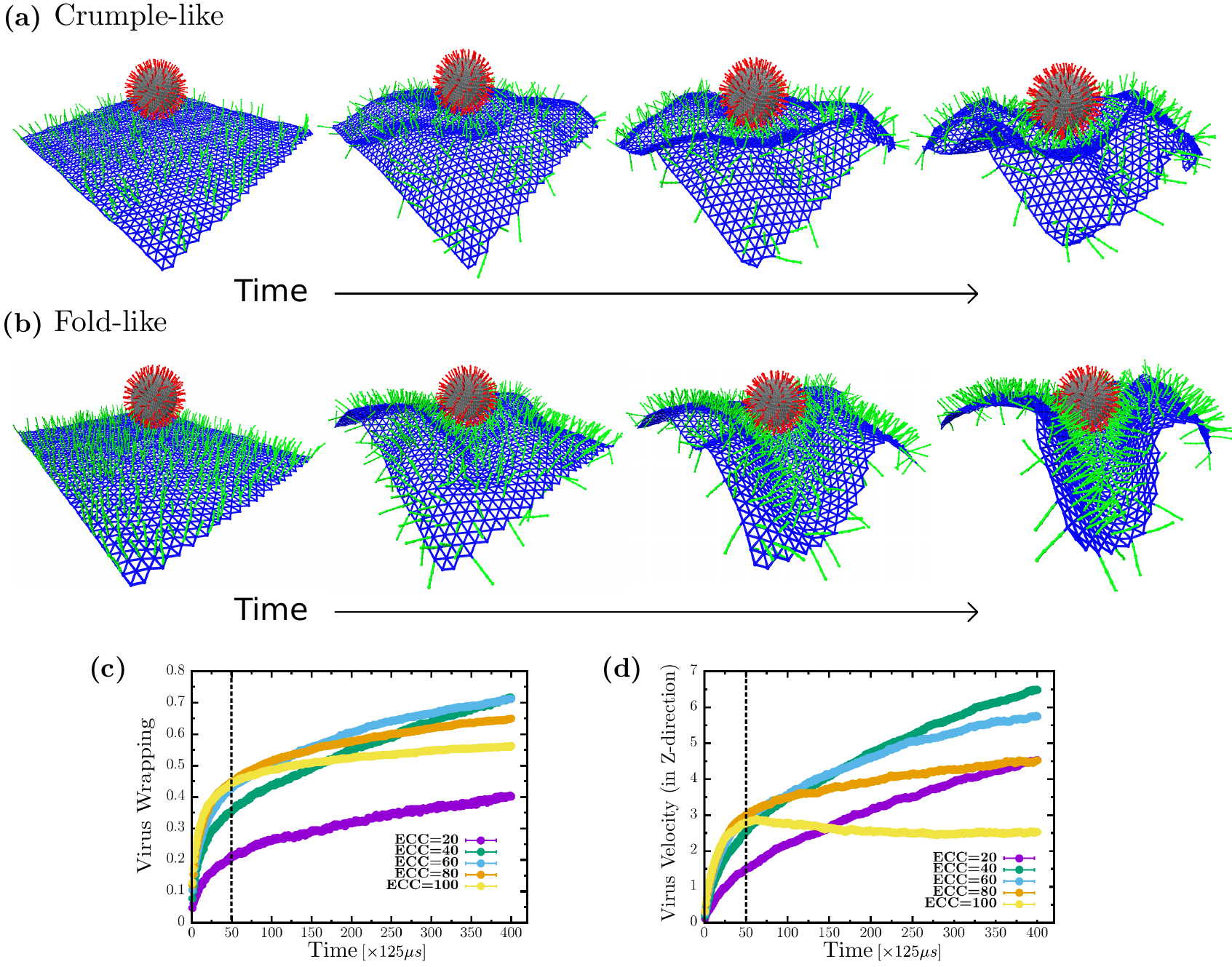}
  \caption{ { \it Cell surface folds wrap the virus faster than cell surface crumples} (a) Cell surface crumpling around the virus while wrapping at low ECC percent coverage (b) Cell surface creates folds while wrapping the virus at optimal ECC percent coverage.  (c) Virus wrapping is fast initially but slows down as fewer ECC are available to attach, eventually has the highest wrapping at optimal ECC percent coverage. (d) Virus velocity in the z-direction shows the highest acceleration with cell surface folds around the virus at optimal ECC percent coverage. The simulation velocity can be converted to a biologically relevant velocity by 200$\mu$m/s. Biologically relevant time units can be obtained by multiplying 125$\mu$s that follow from the time units defined previously. }
\end{figure*}

In Fig3b, we have plotted the cell surface wrapping as a heat map, where the color bar represents the ratio of cell surface near the virus divided by the total cell surface area. At high ECC percent coverage, the cell surface near the virus does not change significantly with cell surface bending rigidity and has less cell surface area near the virus. This is due to effective cell bending rigidity which arises because of volume exclusion among all ECC particles that keep the cell surface shape constant. This effect is more pronounced at high density as more particles lead to higher effective bending rigidity and keeps the surface near-flat. If the cell surface is perfectly flat, the ratio of area under the circle and total surface area is 0.27, which is very near to the value we found for the ratio at a higher percent coverage of ECC. We also observe that for low and mid-ECC densities at zero-bending rigidity, it takes more cell surface area to wrap viruses compared to the non-zero bending rigidity. From Fig. 3a, we see crumples forming at low-mid ECC with zero-bending and folds forming with non-zero-bending rigidity. Therefore, crumples take more cell surface area than folds to wrap the virus. This is more pronounced at the optimal ECC percent coverage of 40 percent, where the difference between cell surface area taken crumple and folds are about 10 percent. While a 10 percent difference is not a substantial difference for the uptake of one virus, it may become more substantive for multiple viruses in addition to the usual material that is endocytosed. 

Furthermore, we plotted the cell surface shape index as a function of ECC percent coverage and cell surface bending rigidity in Fig. 3c. Just as with the dimensionless GI, at low or mid-ECC percent coverage and zero bending rigidities, the larger shape index indicates crumples in red. However, for non-zero bending rigidity, the shape index decreases, as indicated in blue, thereby indicating more efficient wrapping. This is more pronounced at the optimal ECC percent coverage of 40 percent. Note that for the higher ECC percent coverage, the shape index decreases from the crumple value; however, the overall curvature of the cell surface begins to change near the virus to head towards anti-wrapping, if you will.

%Cell surface wrapping define the nearness of cell surface to the virus with non-zero bending rigidity spike-ECC attachment doesnot gurantee that cell surface is also near the virus.

%the observe that we get crumples with zero cell rigidities but get folds for non-zero cell rigidities. We also changed the cell surface coreceptor densities and found the clean folds are forming at 40 percent density for the reasons we expanded upon in previous sections.

%To quantify the folds crumples, and shapes in between, we defined a cell surface wrapping that measures the nearness of the cell surface to the virus. "explain the method." From Fig. 4B, we observe that. 

%Furthermore, we found that the shape index of the cell surface wrapping captures the folds from the crumples.

\subsection{Folds wrap faster than crumples}

We now explore the dynamics of viral wrapping. From Fig. 4a and b, we can see crumple and fold formation, respectively, as the cell surface wraps around the virus. In Fig. 4a, the cell surface is covered with a lower ECC percent coverage, 20 percent leading to crumpling around the virus, and in Fig. 4B, the cell surface has the optimal ECC coverage of 40 percent, where the cell surface is folding towards the virus. We analyzed these configurations by plotting the virus wrapping as a function of simulation time for multiple ECC densities. 

From Fig. 4c, at lower ECC percent coverage, the virus is wrapped slowly, but for higher percent coverage of coreceptors, the cell surface wraps fast initially. To be specific, we observe two regimes of virus wrapping with time, an initial faster wrapping before the dashed line in Fig. 3c, where spikes are attaching to many coreceptors, and a slower regime at later times, after the dashed line, with the virus attached to a lower number of coreceptors leading to slower viral wrapping. In terms of coreceptor percent coverage, virus spikes do not bind many coreceptors to wrap initially for lower coreceptor density. However, as the cell surface crumples, spikes find more coreceptors to bind, increasing virus wrapping with time but eventually achieving low viral wrapping. For higher percent coverage, spikes initially find many coreceptors to bind, but since there is only one-to-one interaction allowed, spikes located at the higher side of the virus struggle to find more later as the shielding effect sets in due to crowding imposed by volume exclusion. Finally, at optimal percent coverage, we see that even though wrapping starts relatively slowly initially, it catches up as more and more coreceptors come into the range of virus spike interaction due to the folding of the cell surface, leading to the highest virus wrapping such that crowding effects are minimized.

We also investigate the virus's engulfing velocity to further understand the role of folds and crumples in viral uptake. We plotted the virus velocity in the z-direction, or perpendicular with respect to the cell surface, see Fig. 4d. Here, we also observe two regimes, an initial fast regime before the dashed line, where virus velocity is similar for all densities except the lowest, as in this region viral spikes are attaching to many coreceptors. We also have a second regime later, after the dashed line, where the virus engulfing velocity is distinct depending on coreceptor percent coverage. For lower percent coverage, the virus engulfing velocity is slower because of the low availabilities of coreceptors, but with higher percent coverage, virus velocity slowly increases with time even though coreceptors are available; due to crowding, additional virus spikes cannot attach to the coreceptors. For the optimal coreceptor percent coverage, we see an increasing velocity over time as spikes keep finding more coreceptors to attach. Thus, the cell surface that folds is faster in catching the virus than cell surfaces that crumple around the virus.

\begin{figure*}[t]
\includegraphics{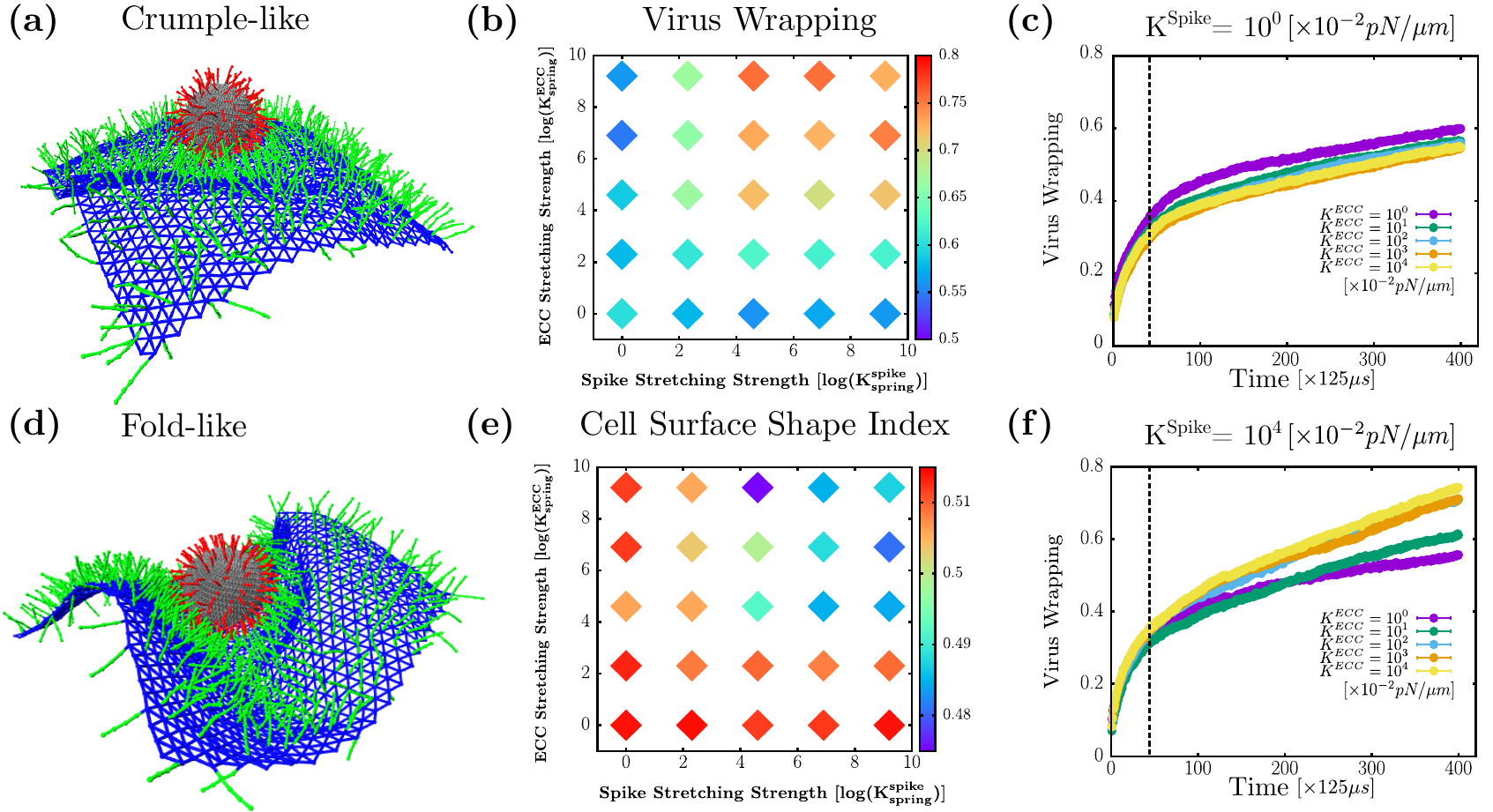}
  \caption{{ \it Changes in ECC and spike stretching strength can lead to crumples (at optimal ECC density).}  (a) Simulation snapshot for $K^{spike}_{spring}=10^{0}$ and $K^{ECC}_{spring}=10^{0}$ show crumpled cell surface (b) log-log plot:  Heat map of virus wrapping showing at low spike and ECC spring strength gives low wrapping compared to high spike and ECC spring strength values (c) and (f) viral wrapping as a function of time for $K^{Spike}_{spring}=10^{0}$ and $K^{Spike}_{spring}=10^{4}$ for various values of $K^{ECC}_{spring}$ (d) Simulation snapshot for $K^{spike}_{spring}=10^{4}$ and $K^{ECC}_{spring}=10^{4}$ showing folded cell surface (e) log-log plot: Heat map of cell surface shape index have a higher value at low streching strength of ECC and spikes indicating crumple formation, but at high ECC and spike strength has lower shape index points to fold formation. Biologically relevant time units can be obtained by multiplying 125$\mu$s that follow from the time units defined previously. For spring constants, the conversion factor is $10^{-2}\,\,pN/nm$.}
\end{figure*}

\subsection{ECC and spike stretching drives the system from folds to crumples}

In this subsection, we investigate the effects of the mechanical properties of spike and ECC on viral wrapping. Indeed, filaments on the cell surface have varied mechanical properties depending on the type of filament. Therefore, we vary the stretching strength of coreceptors and virus spikes on viral uptake. Given this additional mechanical variation, we ask the question, is the optimal percent coverage of coreceptors always guaranteed a fold? To answer this question, we maintain the percent coverage of coreceptors at 40$\%$, which yielded maximal virus wrapping, as shown in Fig. 2c, for 200 spikes (for the given set of mechanical parameters stated previously). From Fig. 5a and 5d, we observe that changing the stretching strength drives the cell surface morphology to crumples, even at the optimal percent coverage of coreceptors. In Fig 5a, coreceptors and spikes have low stretching strength of $10^{-2}pN/nm$. The attraction between spikes and ECC with low stretching cost energy makes them more accessible to each other, leading to many attachments between them. Even though we have a significant number of spikes connected to coreceptors, the cell surface does not form a fold but a crumple. This is because these low- stretching strength coreceptors cannot transfer enough force to the cell surface to make a more coordinated fold. On the other hand, in Fig. 5d, the spike and ECC have the stretching strength of $10^{2} pN/nm$, which is enough to generate forces to fold the cell surface around the virus. Thus, we found that an optimal ECC percent coverage determined for one set of mechanical parameters does not ensure efficiency that virus wrapping can always be achieved. The stretching rigidity of spikes and filaments on the cell surface also constrains the formation of folds.

We vary the stretching strength of spike and coreceptors up to four orders of magnitude given in Table 1. In Fig. 5b, we plot the heat map (on log-log scale) of virus wrapping with respect to the ECC and spike stretching strength. We observe two regions, first a low wrapping region at low spike stretching strength which persists irrespective of coreceptors stretching strength. For low coreceptors stretching strength also, we find low wrapping, irrespective of spike stretching strength, as depicted in blue on the heat map. The second region has high virus wrapping for the high spike and coreceptors stretching strength, as shown in red. Our findings imply that the coreceptor and spike must have high stretching strength to get a high virus wrapping. To identify the folds and crumples qualitatively, we plotted the cell surface shape index in Fig. 5e. We find that for low spike and ECC stretching strength, we measure a higher shape index which is consistent with cell surface crumpling around the virus. At higher spike and ECC stretching strength, we measure a lower value of the cell surface shape index, indicating fold formation. These results are consistent with the previously observed shape index behavior in Fig 4c.

We also examine the time series analysis to understand how the fold and crumple formation mechanism changes due to the stretching strength of spike and coreceptors. In Fig 5c, we set the stretching strength of spikes constant at the lowest explored value of $10^{-2}$pN/nm and vary the coreceptors stretching strength from $10^{-2}$pN/nm to $10^{2}$ pN/nm. We observe that changing coreceptors stretching strength does not contribute much to the virus wrapping behavior as it just increases with time having similar trends for all coreceptor's stretching strength values. On the other hand, from Fig. 5f with a high spike stretching strength of $10^{2}$pN/nm, we see that low coreceptor stretching values are associated with wrapping slowly and get less virus wrapping than for high coreceptor stretching strength. This is because, with the higher stretching strength of the spike and coreceptor, the cell surface folds, leading to the faster wrapping of the virus.

We demonstrate that having optimal coreceptors percent coverage is not enough for efficient wrapping via folds. Coreceptors and spikes also require a threshold stretching strength above which the cell surface folds to achieve more wrapping. We also investigate the effects of the bending rigidity of filamentous ECCs and virus spikes. Bending rigidity does not impact the cell surface morphologies as stretching. See Fig. 2 of the SI. Finally, as for varying viral rigidity, we obtain results consistent with previous work \cite{Shen2019} that softer viruses are harder to wrap than more rigid viruses. See Fig. 3 of the SI.

%Why explore this? what's the biological reason for it?

%Crumple vs fold, no garuntee of folds at optimal density

%Virus and mebrane wrapping shows what and what's the explanation behind this behavriour

%What's the mechanisim at exterme values of stretching strength

%Conclusion of this section

%\subsection{Bending rigidity helps with wrapping}

%\subsection{Soft viruses are more easily wrapped with co-recptors}

\section*{Discussion}

Our study suggests that cells whose surfaces are optimally populated with filamentous protein structures acting as coreceptors are more likely to be infected as they uptake the virus faster and use relatively less cell surface area per individual virus so that more virus-like particles can be uptaken. At the optimal percent coverage, the cell surface makes folds around the virus, and folds are faster and more efficient at wrapping the virus than crumple-like wrapping. Our study also finds that cell surface bending rigidity helps generate folds, as bending rigidity enhances force transmission across the surface. We also conclude that such an optimal percent coverage does not always ensure a fold formation, as changing mechanical parameters, such as the stretching stiffness of the ECC or the virus spikes, can drive the crumple-like formation of the cell surface.

There has been much work exploring the role of viruses or nanoparticle spikes and how their mechanical properties affect endocytosis. However, these studies treat receptors as sticky particles on the cell surface without any degree of freedom \cite{Moreno2022, Lin2020, Li2020c, Shen2019, Xia2017, Ding2012, Li2012}. On the other hand, there have also been many studies of sticky sites embedded in a cell membrane affect endocytosis, though, again, they do not have their own degrees of freedom \cite{Zhang2014, Li2021a}. To our knowledge, this work is the first to consider the physical properties of receptors, including density, stretching, and bending energetic costs on both the cell and viral surface in viral wrapping. In doing so, we find a key quantity to focus on in terms of an effective stiffness of the cell surface that depends on the density of filaments attached and their own intrinsic mechanics. Work will be needed to quantify this property in larger sheets without the virus-like particles. Moreover, it is interesting to determine how material on the outside of a cell, both the virus and the extracellular filaments, can reshape the cell cortex in nontrivial ways in terms of folds versus crumples, which then drives shape changes in the connected plasma cell membrane. Interestingly, earlier work on endocytosis has focused on how the underlying cell cortex can modify the shape of the plasma cell membrane from spherical wrapping to more cylindrical wrapping in yeast \cite{Zhang2015b}. In this work, we unlock a much broader range of shapes for further study.

The earlier experimental finding that extracellular vimentin enhances SARS-CoV-2 uptake is intriguing \cite{Suprewicz2022, Amraei2022}. It turns out that extracellular vimentin helps other viruses and bacteria to enter the cell \cite{Ramos2020}. As there exist vimentin-null mice (but not actin-null nor microtubule-mull mice)~\cite{vimentinnull}, the fact that viruses and bacteria have evolved to interact with vimentin is no surprise when hijacking the machinery of a cell. In doing so, they interact with a higher-order, optimizing construct instead of an essential,  functioning construct to replicate and redeploy without dramatically altering cell function. Presumably, this is yet another evolutionary pressure on viruses to optimize their interaction with vimentin on both the outside and inside of cells. Therefore, our focus on constructs outside and/or attached to the cell emphasizes the importance of the microenvironment of a cell, even for endocytosis. The importance of the tumor microenvironment has now become a cornerstone for understanding cancer, with many modeling efforts underway to make predictions \cite{DiVirgilio2018, Lim2018, DePalma2017, Parker2020}. Given the work here, we now argue that the notion of microenvironment has an impact on viral and bacterial infections and will contribute towards our understanding of the variability of health impacts of such infections. While our inspiration here has been extracellular vimentin, glycolipids that bind the protein lectin to form a filamentous-like complex emanating from the cell surface and can play a role in clathrin-independent endocytosis~\cite{lectin}. Our results point to a potential mechanical role for this complex in endocytosis.

We have assumed that the virus-like particle is already attached via a small receptor to the cell surface and quantified viral wrapping. However, it would also be interesting to study how the cellular microenvironment affects the trajectory of a nanoparticle to find that initial receptor in terms of search strategy \cite{Marbach2022}. We have also assumed a one-to-one interaction between ECC and spike without any kinetics, i.e., no attaching or detaching rate is considered in this model. Finally, as our cell surface is a deformable sheet to which filamentous structures attach, we will investigate how the nanoparticle enters the cell via a pinch-off mechanism by extending our work to include a cellular fluid membrane. In the earlier work quantifying cylindrical endocytosis in yeast, the proposed pinch-off mechanism is a Pearling instability driven by BAR proteins acting on both the cortex and the plasma cell membrane \cite{Zhang2015b}. As additional morphologies of the cell surface are proposed in mammalian cells \cite{Abouelezz2022, Jin2022, Lanzetti2001}, perhaps a Pearling instability or additional mechanisms will be discovered. To make such predictions, the richness of biology must be reflected in analytical or computational modeling in at least some minimal manner.

%eplace the deformable viral shell with a fluid shell 
%Our model does not take account of the fluid lipid membrane as we were interested to understand the ECC's role in viral uptake, which is joined to part of the actomyosin layer under the cell membrane Unlike other work \cite{Moreno2022}, we only cosider viruses that could have spike around 200. For the next iteration of the model, we would consider these factors and investigate the virus search strategies. However, in studying wrapping in the context of deformable sheet, we are able to study crumples versus folds to obtain a more intricate picture of how the elasticity (on short time scales) can drive shape changes in the cell membrane.  Ruffles.....Effective bending rigidity.....

%Conclude with paragraph on viral evolutionary pressures in terms of a cell's own ``microenvironment'' or what exists outside the cell and talk about vimentin and viral infection more generally.....

%Other models in the literture \cite{Moreno2022}

%CITE desarno paper

%Limitations
%No fluid membrane
%virus is already tethered
%one-one interaction

%Future directions
%Viruses may have different search strategies that are not explored here \cite{Boulant2015}
%Add kon and koff rate for ECC-spike.
%Multi-body interaction is allowed

\section*{Acknowledgements}
AEP and JMS acknowledge an NSF RAPID grant 2032861. AEP acknowledges NIH R35 GM142963. CDS acknowledges NSF DMR 2217543. SG acknowledges a dissertation graduate fellowship from Syracuse University. The authors also acknowledge the Syracuse University HTC Campus Grid and NSF award ACI-1341006 for providing computing resources.

\bibliography{library2}

%\section{Simulation results}

%\begin{Figure*}[t]
%\includegraphics{SI_3.pdf}
 % \caption{Simulation snapshots of time series showing change in wrapping due to streching rigidity Top: Spike and co-receptors's streching rigidity is 1, showing a higher connections doesn't lead to higher wrapping, Bottom: Spike and co-receptors's has relativly high spring strength demonstarting that it is required for wrapping of virus. }
%\end{Figure*}

\pagebreak

\onecolumngrid

\setcounter{equation}{0}
\setcounter{figure}{0}
\setcounter{table}{0}
\setcounter{page}{1}
\makeatletter
\renewcommand{\theequation}{S\arabic{equation}}
\renewcommand{\thefigure}{S\arabic{figure}}

\section{Supplementary Figures}
\begin{figure*}[hbt!]
\includegraphics{  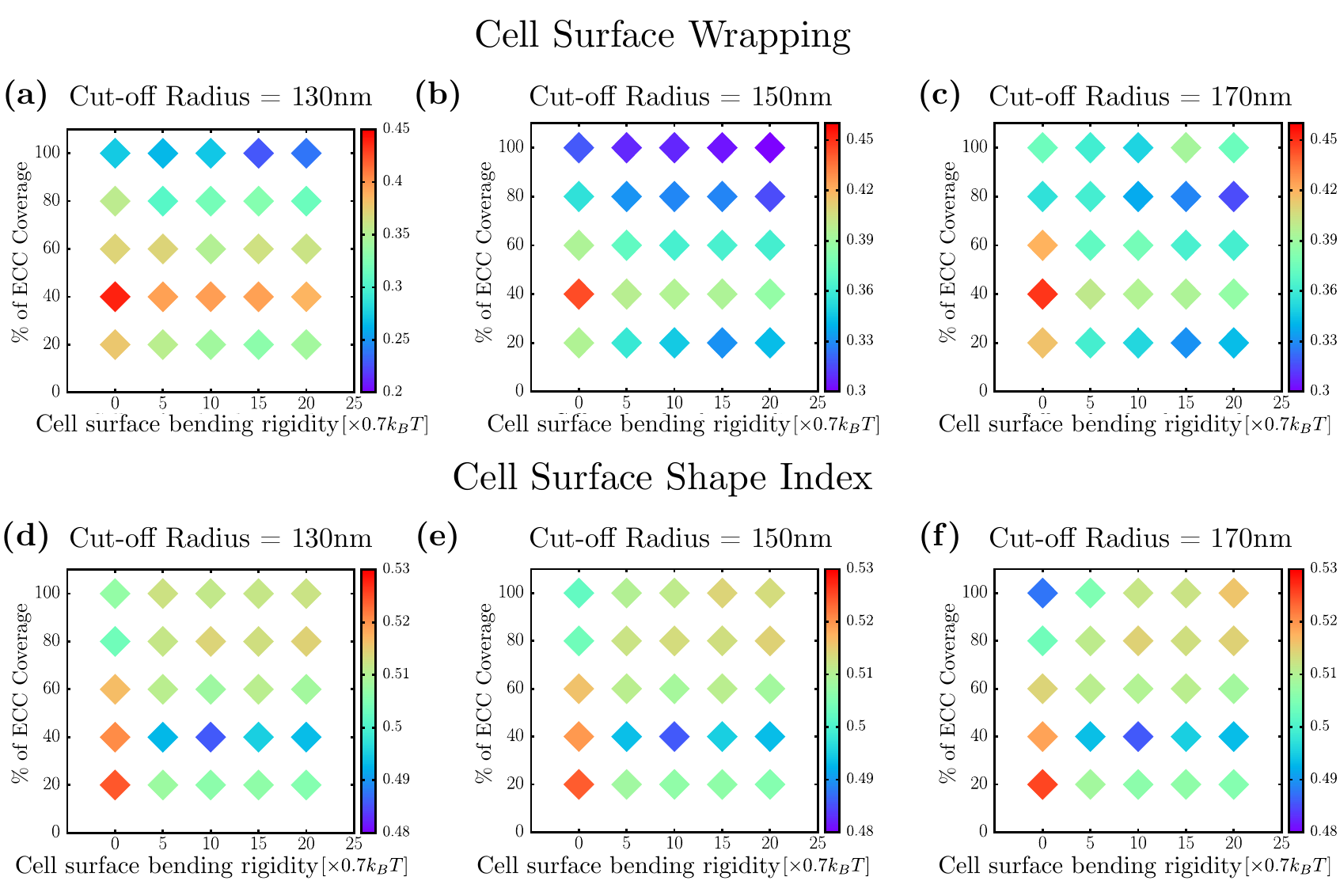}
\caption{ { Cell surface wrapping and shape index is robust with the varying cylinder's radius cut-off} }  
 \end{figure*}

\begin{figure*}[hbt!]
\includegraphics{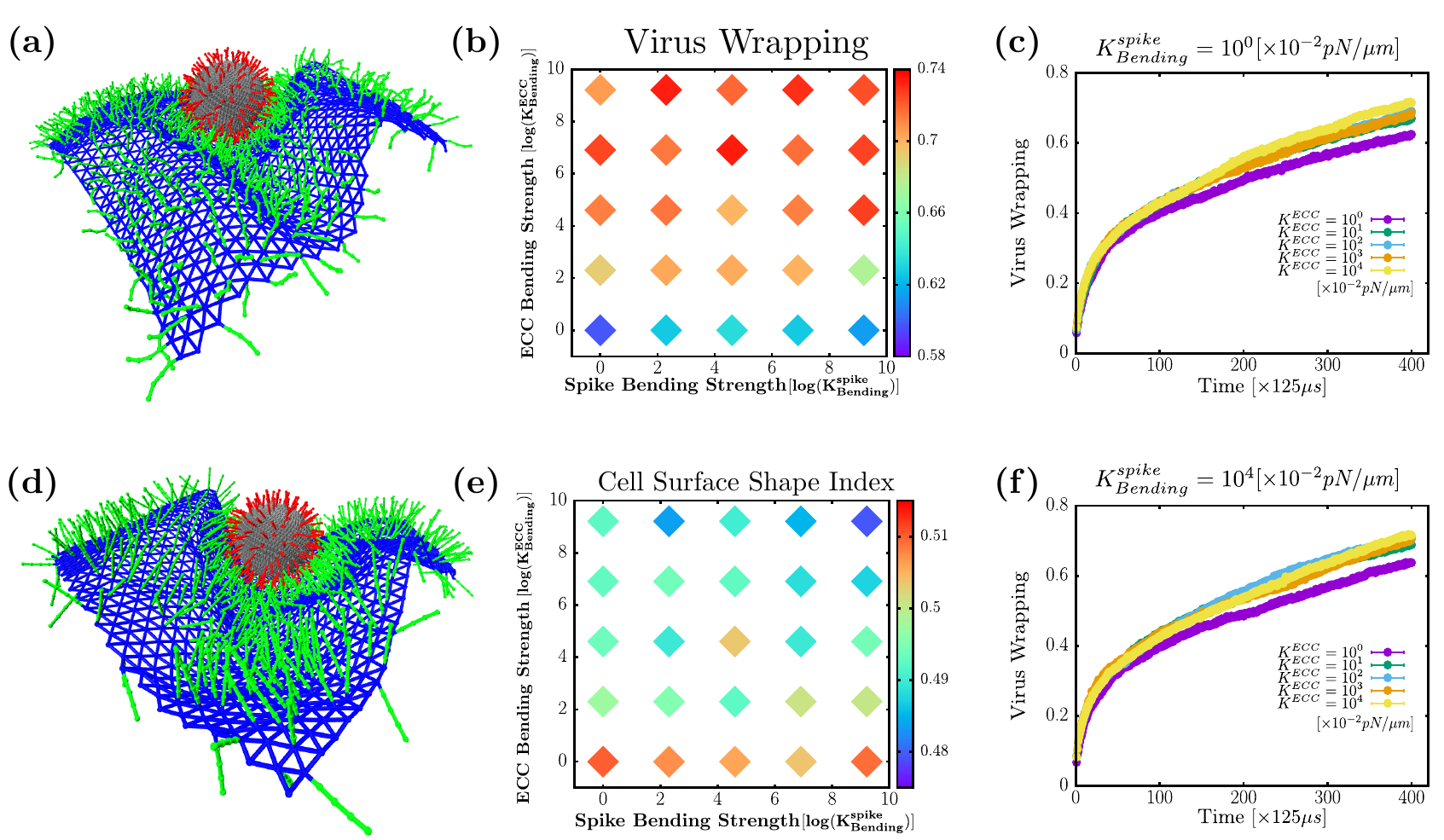}
\caption{ {\it Varying ECC and spike bending rigidities does not lead to cell surface crumples} (a) Simulation snapshot for $K^{spike}_{Bending}=10^{0}$ and $K^{ECC}_{Bending}=10^{0}$ show folded state  (b) log-log plot:  Heat map of virus wrapping showing at low ECC bending strength irrespctive of spike bending strength gives low wrapping compared to high ECC bending strength (c) and (f) Virus wrapping as a function of time for $K^{spike}_{Bending}=10^{0}$ and $K^{spike}_{Bending}=10^{4}$ for various values of $K^{ECC}_{Bending}$ (d) Simulation snapshot for $K^{spike}_{Bending}=10^{4}$ and $K^{ECC}_{Bending}=10^{4}$ showing sharply folded cell surface (e) log-log plot: Heat map of cell surface shape index have a higher value at low bending strength of spike indicating soft folds formation, but at high ECC bending strength has lower shape index points to sharp fold formation. Biologically relevant time units can be obtained by multiplying 125$\mu$s that follow from the time units defined previously.}  
 \end{figure*}

\begin{figure*}[hbt!]
\includegraphics{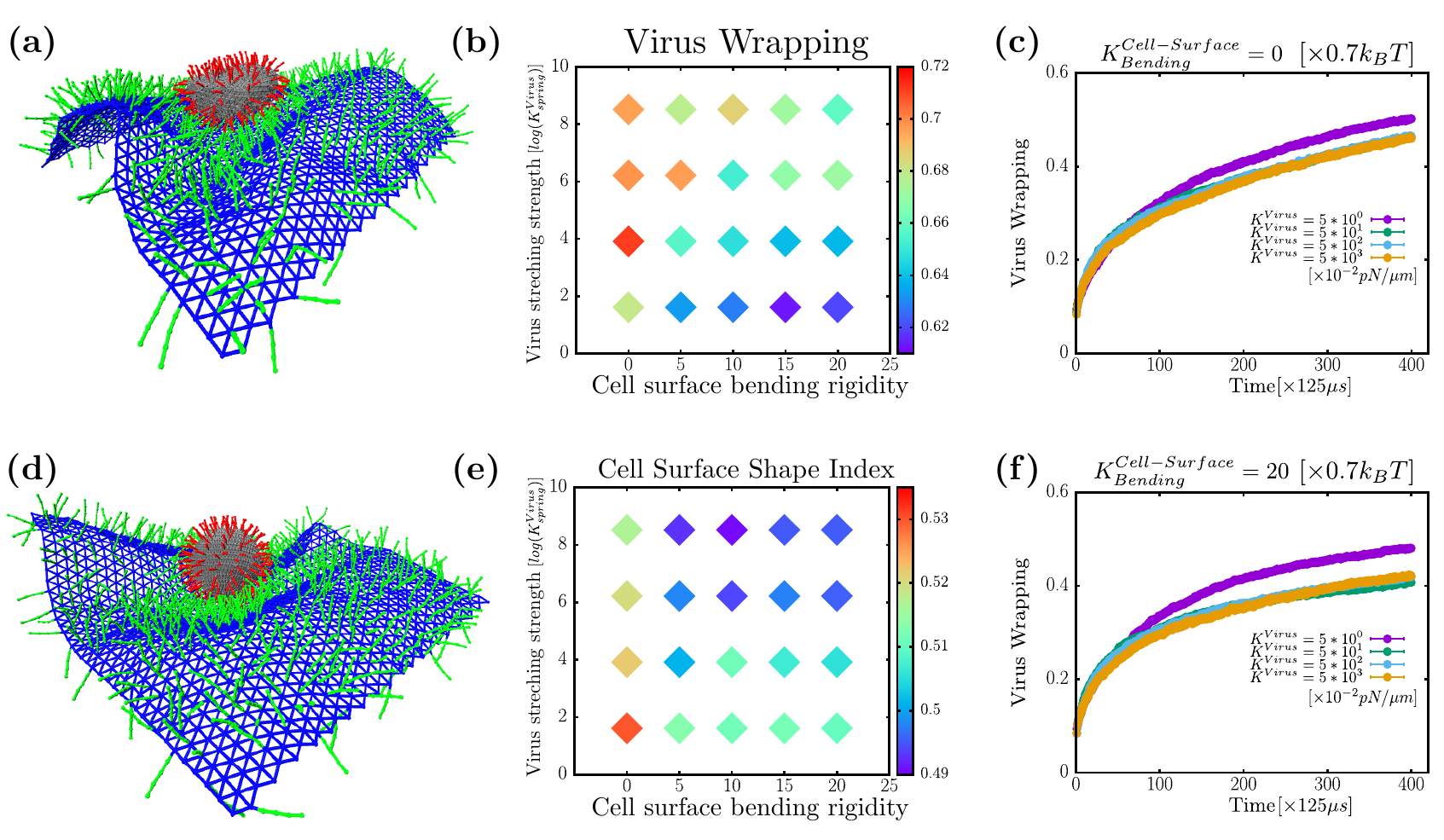}
\caption{ {\it Varying virus stretching rigidity} (a) Simulation snapshot for $K^{Virus}_{Spring}=5*10^{0}$  (b) log-log plot:  Heat map of virus wrapping as a function of virus stretching strength and cell surface bending rigidity. Since the virus is deformable, virus wrapping here is defined as the ratio of occupied spikes divided by the total number of virus spikes.  (c) and (f) Virus wrapping as a function of time for $K^{Cell-Surface}_{Bending} = 0$ and $K^{Cell-Surface}_{Bending} = 20$ for various values of $K^{Virus}_{Spring}$ (d) Simulation snapshot for $K^{Virus}_{Spring}=5*10^{4}$ (e) log-log plot: Heat map of cell surface shape index as a function of virus stretching strength and cell surface bending rigidity. Biologically relevant time units can be obtained by multiplying 125$\mu$s that follow from the time units defined previously.}   
 \end{figure*}

\end{document}